%% file: main.tex
\documentclass[conference]{IEEEtran}

\input{head}
\input{include}

\begin{document}

\title{Concurrent and Adaptive Extreme Scale Binding Free Energy
Calculations}

\author{
	\IEEEauthorblockN{
		Jumana Dakka\IEEEauthorrefmark{1}, 
		Kristof Farkas-Pall\IEEEauthorrefmark{2}, 
		Matteo Turilli\IEEEauthorrefmark{1}, \\ 
		David W. Wright\IEEEauthorrefmark{2},  
		Peter V. Coveney\IEEEauthorrefmark{2}, 
		Shantenu Jha\IEEEauthorrefmark{1}\IEEEauthorrefmark{3}
	}
	\IEEEauthorblockA{
		\IEEEauthorrefmark{1}Rutgers University, Piscataway, NJ, USA
	}
	\IEEEauthorblockA{
		\IEEEauthorrefmark{2}Center for Computational Sciences, University College London, UK
	}
	\IEEEauthorblockA{
		\IEEEauthorrefmark{3}Brookhaven National Laboratory, Upton, NY, USA
	}
}

\maketitle

\begin{abstract}
The efficacy of drug treatments depends on how tightly small molecules bind to
their target proteins. The rapid and accurate quantification of the strength
of these interactions (as measured by `binding affinity’) is a grand challenge
of computational chemistry, surmounting which could revolutionize drug design
and provide the platform for patient-specific medicine. Recent evidence
suggests that molecular dynamics (MD) can achieve useful predictive accuracy
($\leq$ 1 kcal/mol). For this predictive accuracy to impact clinical decision
making, binding free energy computational campaigns must provide results
rapidly and without loss of accuracy. This demands advances in algorithms,
scalable software systems, and efficient utilization of supercomputing
resources. We introduce a framework called HTBAC, designed to support accurate
and scalable drug binding affinity calculations, while marshaling large
simulation campaigns. We show that HTBAC supports the specification and
execution of free-energy protocols at scale. This paper makes three main
contributions: (1) shows the importance of adaptive execution for 
ensemble-based free energy protocols to improve binding affinity accuracy; (2)
presents and characterizes HTBAC -- a software system that enables the scalable
and adaptive execution of binding affinity protocols at scale; and (3) for a
widely used free-energy protocol (TIES), shows improvements in the accuracy of
simulations for a fixed amount of resource, or reduced resource consumption
for a fixed accuracy as a consequence of adaptive execution.

\end{abstract}

\section{Introduction}\label{sec:intro}
\input{introduction}

\section{Background}\label{sec:related-work}
\input{related_work2}

\section{Science Drivers}\label{sec:science-drivers}
\input{fe_protocols}
\input{physical_systems}
\section{High-Throughput Binding Affinity Calculator
(HTBAC)}\label{sec:htbac}
\input{design_implementation}

\section{Experiments}\label{sec:experiments}
\input{experiments}

\section{Discussion and Conclusion}\label{sec:discussion}
\input{conclusions}

\section*{Software and Data}
All experimental data can be found at
\url{https://github.com/radical-experiments/htbac-escience-18}. HTBAC (MIT
license) can be found at: \url{https://github.com/radical-cybertools/htbac}

\section*{Acknowledgments}
JD acknowledges Dept. of Education Award P200A150217 for her
fellowship. PVC acknowledges the support of the EU H2020 CompBioMed (675451),
EUDAT2020 (654065) and ComPat (671564) projects. Access to Blue Waters was
made possible by NSF 1713749. The software capabilities were supported by
RADICAL-Cybertools (NSF 1440677). We thank Vivek Balasubramanian and Andre
Merzky for support with RADICAL-Cybertools.

\bibliographystyle{IEEEtran}
\bibliography{rutgers,ucl}

\end{document}

%% file: head.tex
\usepackage{booktabs}
\usepackage[utf8]{inputenc}
\usepackage{keyval}
\usepackage{listings}
\usepackage{moresize}
\usepackage{multirow}
\usepackage{paralist}
\usepackage{rotating}
\usepackage{soul}
\usepackage{srcltx}
\usepackage{url}
\usepackage[dvipsnames]{xcolor}
\usepackage{xspace}
\usepackage{wrapfig}
\usepackage{enumitem}

\definecolor{listinggray}{gray}{0.95}
\definecolor{darkgray}{gray}{0.7}
\definecolor{commentgreen}{rgb}{0, 0.4, 0}
\definecolor{darkblue}{rgb}{0, 0, 0.6}
\definecolor{purple}{rgb}{0.6, 0, 0.6}
\definecolor{middleblue}{rgb}{0, 0, 0.75}
\definecolor{darkred}{rgb}{0.4, 0, 0}
\definecolor{brown}{rgb}{0.5, 0.5, 0}
\definecolor{dkgreen}{rgb}{0,0.5,0}
\definecolor{orange}{rgb}{1,.5,0}
\definecolor{dandelion}{cmyk}{0,0.29,0.84,0}

\usepackage[normalem]{ulem}
\makeatletter
\def\cyanuwave{\bgroup \markoverwith{\lower3.5\p@\hbox{\sixly \textcolor{cyan}{\char58}}}\ULon}
\def\reduwave{\bgroup \markoverwith{\lower3.5\p@\hbox{\sixly \textcolor{red}{\char58}}}\ULon}
\def\blueuwave{\bgroup \markoverwith{\lower3.5\p@\hbox{\sixly \textcolor{blue}{\char58}}}\ULon}
\font\sixly=lasy6 
\makeatother

\usepackage{pgfplots}
\pgfplotsset{compat=newest}
\usepgfplotslibrary{fillbetween}
\usetikzlibrary{patterns}

\usepackage{siunitx}
\DeclareSIUnit{\calorie}{cal}
\usepackage[outline]{contour}

\usepackage{makecell}

%% file: include.tex
\newcommand{\B}[1]{\textbf{#1}\xspace}

\newcommand{\up}{\vspace*{-0.5em}}



%% file: introduction.tex
Drug discovery and design is immensely expensive with one study putting the
current cost of each new therapeutic molecule that reaches the clinic at
US\$1.8 billion~\cite{Paul2010}. A diversity of computational approaches,
specifically binding free energy calculations which rely on physics-based
molecular dynamics simulations (MD) have been developed~\cite{Mobley2012}
and blind tests show that many have considerable predictive
potential~\cite{Mey2017,Yin2017}. The development of commercial approaches
that claim accuracy of below 1 kcal mol$^{-1}$~\cite{Wang2015} has led to
increased interest from the pharmaceutical industry~\cite{Ganesan2017} in
designing computational drug campaigns.

These improvements can be attributed to many advances in methodologies and
hardware. Specifically, ensemble-based binding free energy calculations,
which favor many shorter simulation trajectories over few longer simulations,
have been shown to increase sampling efficiency whilst also reducing time to
insight~\cite{weis_ligand_2006}. For binding affinity calculations to gain
traction, they must have well-defined uncertainty and consistently produce
statistically meaningful results.

Computational drug campaigns rely on rapid screening of millions of
compounds, which start with an initial screening of candidate compounds to
filter out the ineffective binders before using more sensitive methods to
refine the structure of promising candidates. Two prominent ensemble-based
free energy protocols, ESMACS and TIES~\cite{Bhati2017}, have shown the
ability to consistently filter and refine the drug design process. The ESMACS
(enhanced sampling of molecular dynamics with approximation of continuum
solvent) protocol provides an ``approximate'' endpoint method used to screen
out poor binders. The TIES protocol (thermodynamic integration with enhanced
sampling) uses the more rigorous ``alchemical'' thermodynamic integration
approach as implemented in NAMD \cite{Phillips2005, Straatsma1991}. These
protocols have produced statistically meaningful results for industrial
computational drug campaign~\cite{Wan2017brd4}.

In recent years, considerable effort has been put into improving the
efficiency of free energy calculations~\cite{Klimovich2015, Naden2016,
Kaus2013}. As drug screening campaigns can cover millions of compounds and
require hundreds of millions of core-hours, it is important that these
calculations be effective and aim to optimize the accuracy and precision of
results. This is challenging as, by definition, drug discovery involves
screening compounds that are highly varied and potentially unique in their
chemical properties. The variability in the drug candidate chemistry results
in diverse sampling behavior that contributes to the statistical uncertainty
of binding free energy predictions. Further, a particular difficulty comes
from the fact that not all changes induced in protein shape or behavior are
local to the drug binding site and, in some cases, simulation protocols will
need to adjust to account for complex interactions between drugs and their
targets within individual studies.

Traditionally, the simulated duration of free energy calculation are
conservatively determined to account for likely slowest convergence and worst
case scenarios. This approach has at least two shortcomings: it potentially
wastes valuable computational resources  and fails to account for the
different value of the simulations results. For example, in a drug campaign it
is more important to understand how strong is the binding of the best compound
candidates than precisely know how weak is the interactions of the worst
compound.


Key to successful campaigns is identifying when small chemical changes result
in large binding strength changes. This can mean that the parameters which are
important to campaigns evolve as the study progresses. Here we show how
adaptive approaches using ensemble-based free energy protocols can be designed
to capture unique chemical properties and customize the simulations for a
candidate to make the most effective use of computational resources, thereby
improving statistical uncertainties.

Adaptive approaches on high performance computers (HPC) require software
systems that make runtime decisions based on intermediate
results~\cite{cosb18,adaptivebiomolecular}. To achieve scalability and
efficiency, these software systems must also support efficient dynamic
resource allocation. Further, such adaptivity cannot be performed via user
intervention and hence automation of the control logic and execution is
important. We have developed the High-Throughput Binding Affinity Calculator
(HTBAC) to enable the scalable execution of adaptive algorithms.

This paper makes three main contributions: (1) shows the importance of
adaptive execution for ensemble-based free energy protocols to improve binding
affinity accuracy; (2) presents and characterizes HTBAC, a software system
that enables the scalable and adaptive execution of binding affinity protocols
at scale; and (3) for a widely used free-energy protocol (TIES), shows
improvements in the accuracy of simulations for a fixed amount of resource, or
reduced resource consumption for a fixed accuracy as a consequence of adaptive
execution.


This paper is organized as follows: Section~\ref{sec:science-drivers}
introduces ESMACS and TIES, two ensemble-based free energy protocols, arguing
how implementing adaptive methodology within TIES could achieve higher
precision with limited resources. Section~\ref{sec:related-work} describes
the motivation for ensemble-based approaches and existing solutions alongside
the limitations in their ability to support adaptive methods.
Section~\ref{sec:htbac} describes the design and implementation of HTBAC and
how we used HTBAC to implement an adaptive and nonadaptive version of TIES.
In Section~\ref{sec:experiments}, we present experiments we performed with
HTBAC to characterize its scalability and overheads, and showing that given a
fixed amount of computing resources, we can achieve better accuracy and
better time to solution using adaptive methods.

%% file: related_work2.tex
Free-energy calculations using MD simulations occur in a wide range of
research including protein folding and assessing small molecule binding.
Free-energy calculations require three main components: (1) suitable
Hamiltonian model; (2) sampling protocol; and (3) estimator of free energy.
Several approaches to computing binding free energies exist, amongst which
relative binding free energy (or binding affinity) calculations are
generating accurate predictions, delivering considerable promise for
computational drug campaigns~\cite{Karplus2005}.

Ensemble-based simulations have been shown to reduce the sampling time
required to deliver the precision necessary to meet the requirements of drug
design campaigns. Several ensemble-based methods are widely used to compute
binding free energies, studying different problem spaces. For example, a
popular approach is to use Markov state models to learn a simplified
representation of the explored phase space and to choose which regions should
be further sampled~\cite{Bowman2010}. Replica exchange with solute tempering
use the Metropolis-Hastings criteria to make periodic decisions about what
regions of the phase space to sample~\cite{Earl2005,Hritz2008,Kim2012}. In
expanded ensemble simulations, thermodynamic states are explored via a biased
random walk in state space~\cite{Lyubartsev1992}. Approaches that learn by
exchanging information have been found to improve sampling results and
decorrelate as fast or faster than standard simulations.

In binding affinity calculations sampling is performed at discrete regions
along the transformation between the two compounds. The choice of where
exactly this sampling occurs is a key determinant of the uncertainty in and
accuracy of the calculations~\cite{Ruiter2013,Ruiter2016}. Increasing
simulations in regions of most rapid change reduces errors on the predicated
binding affinity.

Using ensemble-based methods to compute binding affinities of a large number
of drug candidates involves a hierarchy of computational processes: at the
lowest level is the specific molecular dynamics (MD) simulation using an MD
engine, such as Gromacs or AMBER. An ensemble-based \textbf{algorithm} (or
equivalently \textbf{protocol}) is comprised of multiple such MD simulations
that are collectively used to compute the binding free energy of a single drug
candidate. There are multiple protocols that can be used, each comes with its
specific trade-offs. For example, TIES and ESMACS are two protocols to compute
binding affinities that differ in their accuracy but also their computational
cost. The computational instance implementing a protocol with specific
parameter values, number of simulations and other computational aspects of
that protocol, constitutes a \textbf{workflow}. A workflow may be fully
specified a priori, or it may adapt one or more of its properties, say parameters,
as a consequence of intermediate results. Typically, there is a one-to-many
relationship between protocols and workflows and different workflows can be
used to compute a given binding affinity calculation for a given drug
candidate.

When multiple drug candidates need to be evaluated with certain constraints
and a defined objective, the entire computational activity (i.e., computing
binding affinities for multiple drug candidates) is referred to as a
\textbf{computational campaign}. The objective of the computational campaign
of relevance to this paper is to maximize the number of drug candidates for
which the binding affinity of each individual candidate is determined to
within a (given) acceptable level of error. The campaign is constrained by
the computational resources available, measured in thousand of core-hours. To
meet this objective, each workflow computing the binding affinity of a drug
candidate is adaptively executed.

Executing scalable and adaptive simulation methods on production-grade HPC
resources using ensemble-based methods presents several
challenges~\cite{cosb18,adaptivebiomolecular}. HTBAC addresses these
challenges in the context of a computational campaign to compute
ensemble-based free energy methods.

%% file: fe_protocols.tex
In this section we provide details about ESMACS and TIES specifications and
about adaptive methodologies using TIES\@. We conclude with a description and
validation of the physical systems used in this work.

\subsection{ESMACS and TIES}\label{ssec:esm_ties}

ESMACS and TIES~\cite{Wan2017brd4, Bhati2017} are two free energy calculation
protocols that implement absolute and relative methods, respectively.
Absolute free energy methods calculate the binding affinity of a
\emph{single} drug molecule to a protein, while relative methods calculate
the \emph{difference} in binding affinity between two (usually similar in
structure) drug molecules. Both protocols are designed to use an ensemble MD
simulation approach to enhance the reproducibility and accuracy of standard
free energy calculation techniques (MMPBSA~\cite{Massova1999} in the case of
ESMACS and thermodynamic integration~\cite{Straatsma1988, Straatsma1991} in
TIES). The use of ensemble averaging allows tight control of error bounds in
the resulting free energy estimates.

ESMACS and TIES consists of three main steps: minimization, equilibration and
production MD (in its current implementation all MD steps are conducted in
NAMD~\cite{Phillips2005}). In practice, the equilibration phase is broken
into multiple steps to ensure that the size of the simulation box does not
alter too much over the simulation. During these steps, positional
constraints are gradually released from the structure and the physical system
is heated to a physiologically realistic temperature.

\begin{figure}
  \centering
  \includegraphics[width=\columnwidth]{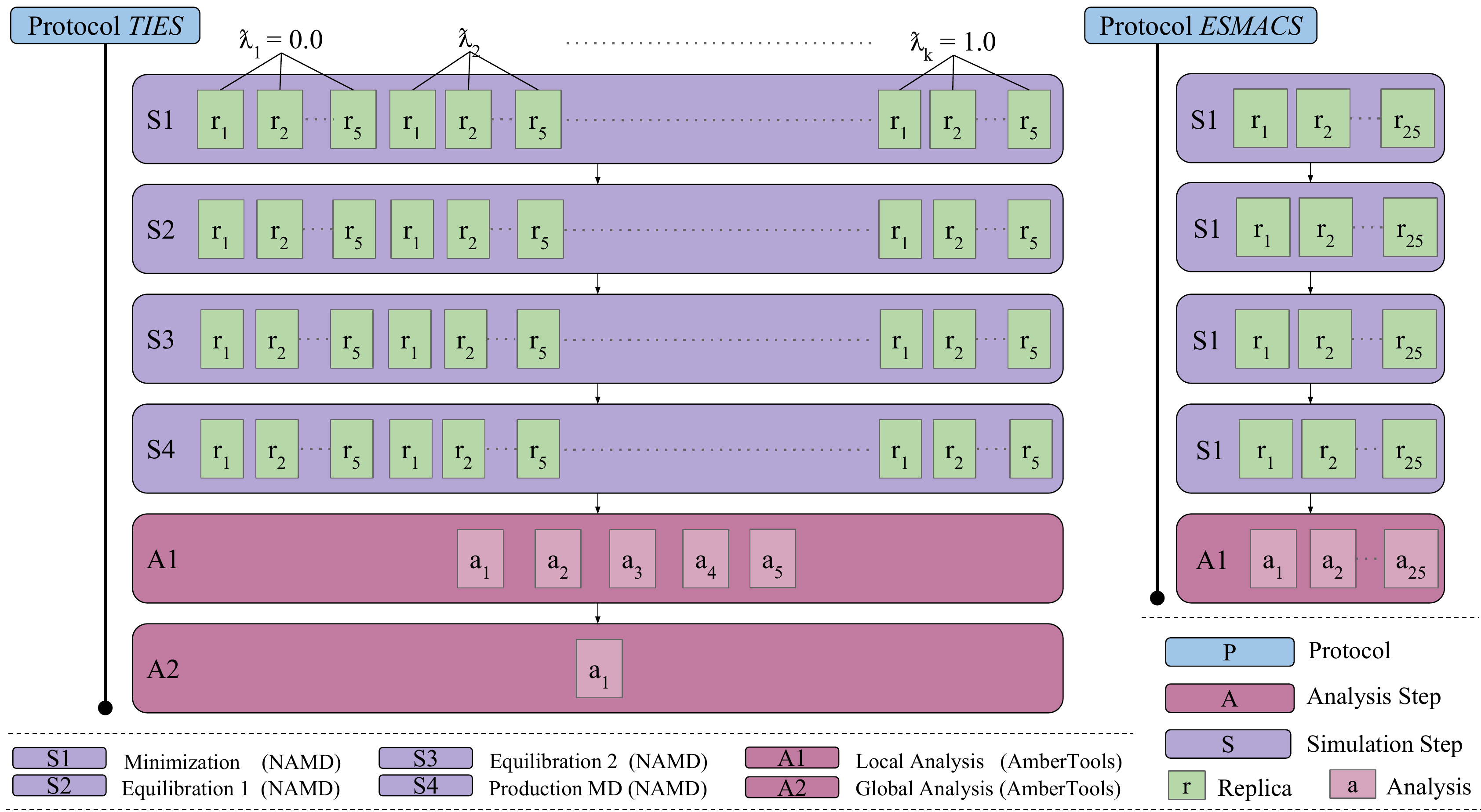}
  \caption{TIES and ESMACS protocols consist of simulations steps followed by
  analysis step(s). ESMACS contains 25 replicas per simulation step; TIES
  contains 5 replicas per $\lambda$ window. We model TIES with 13 $\lambda$
  windows, spawning 65 replicas in each simulation step. All replicas
  simulate 6ns.}\label{fig:ties_esmacs_application}
\up{}
\up{}
\end{figure}

Whilst both protocols share a common sequence of steps, the make-up of the
ensemble is different. In ESMACS, an ensemble consists of a set of 25
\textbf{replicas}, i.e., identical simulations differing only in the initial
velocities assigned to each atom. In TIES, the ensemble contains a set of
\textbf{$\lambda$} windows, each spawning a set of replicas. As a
transformation parameter $\lambda$ increases from 0 to 1, the system
description transforms from containing an initial drug to a target compound
via a series of hybrid states. Sampling along $\lambda$ is then required to
compute the difference in binding free energy. In previous studies, TIES has
been deployed using 65 replicas, evenly distributed among 13 $\lambda$
windows. Following the completion of the simulation steps, both protocols
require the execution of free energy analysis steps. The detailed composition
of ESMACS and TIES protocols is shown in
Fig.~\ref{fig:ties_esmacs_application}.

\subsection{The Value of Adaptivity}\label{ssec:adapt_ties}

The main driver for adaptivity is that computational campaigns will typically
involve compounds with a wide range of chemical properties which can impact
the time to convergence and the type of sampling required to gain accurate
results. There may be cases where it is important to increase the sampling of
phase space, possibly through expanding the ensemble. In general, there is no
way to know exactly which calculation setup a particular system requires
before runtime.

Another driver of adaptivity is that, on occasion, alchemical methods may
converge very slowly. This means that the most effective way to gain accurate
and precise free energy results on industrially or clinically relevant
timescales is to adapt either the workflow corresponding to a specific
protocol or adapt different workflows in relation to each other. The latter
is referred to as \textbf{inter-protocol} adaptivity; the former as
\textbf{intra-protocol} wherein, for example, the parameter values associated
with a specific protocol might change. With thousands of workflows
(corresponding to a protocol instances) to adapt in different ways, this has
the potential to allow for significant optimization.

In TIES, the change in free energy associated with the transformation is
calculated using an adaptive quadrature function which numerically integrates
the values of the $\partial U/\partial\lambda$ across the full set of
simulated $\lambda$ windows. Obtaining accurate and precise results from TIES
using adaptive quadratures requires that the $\lambda$ windows correctly
capture the changes of $\partial U/\partial\lambda$ over the transformation.
This behavior is highly sensitive to the chemical details of the compounds
being studied and varies considerably among candidates. Typically, $\lambda$
windows are evenly spaced between 0 and 1 with the spacing between them set
before execution at a distance determined by the simulator to be sufficient
for a wide range of systems.

However, the number or the location of the $\lambda$ windows that will most
impact the calculation are not known \textit{a priori}, and varies across
candidates. As each window requires multiple simulations, sampling with a
high frequency is expensive. Approximations using evenly spaced $\lambda$
windows reach an acceptable accuracy threshold but adaptive placement of
$\lambda$ windows is likely to better capture the shape of the $\partial
U/\partial\lambda$ curve, leading to more accurate and precise results for a
comparable computational cost.

In this work, we focus on intra-protocol adaptivity which relies on
intermediate runtime results \textit{within} a protocol instance to define
the following set of simulations. Instead of approximating the placement of
all the $\lambda$ windows prior to execution, we run TIES with less $\lambda$
windows and shorter bursts of simulations, analyzing intermediate runtime
results (i.e., trajectories) to seed new and ideally placed $\lambda$
windows.

%% file: physical_systems.tex
\subsection{Physical system description}

Scientific and computational improvements require validation across a number
of protein ligand complexes. We selected 4 proteins and 8 ligands or ligand
pairs to run adaptive free energy calculations. The proteins are the Protein
tyrosine phosphatase 1B (PTP1B), the Induced myeloid leukemia cell
differentiation protein (MC1), tyrosine kinase 2 (TYK2) and the
bromodomain-containing protein 4 (BRD4). Four ligands are alchemical
transformations from one to another (used in TIES), four are single ligands
suitable for absolute free energy calculations (used in ESMACS). All systems
were taken from previously published studies~\cite{Bhati2017}.

Simulations were set up using our automated tool, BAC~\cite{Sadiq2008}. This
process includes parametrization of the compounds, solvation of the
complexes, electrostatic neutralization of the systems by adding counterions
and generation of configurations files for the simulations. The AMBER
ff99SB-ILDN~\cite{Lindorff-Larsen2010} force field was used for the proteins,
and TIP3P was used for water molecules. Compound parameters were produced
using the general AMBER force field (GAFF)~\cite{Wang2004} with Gaussian
03~\cite{Frisch} to optimize compound geometries and to determine
electrostatic potentials at the Hartree–Fock level (with 6-31G** basis
functions). The restrained electrostatic potential (RESP) module in the AMBER
package~\cite{Case2005} was used to calculate the partial atomic charges for
the compounds. All systems were solvated in orthorhombic water boxes with a
minimum extension from the protein of 14 \AA\, resulting in systems with
approximately 40,000 atoms.

%% file: design_implementation.tex
HTBAC is a software system for running ensemble-based free energy protocols
adaptively and at scale on HPC resources. Currently, HTBAC supports protocols
composed of an arbitrary number of analysis and simulation steps, and relies
on the ensemble management system and runtime system provided by the
RADICAL-Cybertools (RCT). HTBAC is designed to be extended to support more
types of protocols and alternative runtime middleware.

\subsection{Requirements}

HTBAC satisfies three main requirements: (1) enable the scalable execution of
concurrent free energy protocols; (2) abstract protocol specification from
execution and resource management; and, (3) enable adaptive execution of
protocols.

Computational drug campaigns increasingly depend on scalable ensemble-based
protocols. This poses at least two major computational challenges. First,
ensemble-based protocols require execution coordination and resource
management among ensemble members, within protocols as well as across
protocols. Second, the setup of execution environments and data management
has to preserve efficient resource utilization. These challenges need to be
addressed by HTBAC as well as the underlying ensemble management and runtime
system.

Adaptive execution of protocols require the ability to change the control
logic of the ensemble execution, based on intermediate results of the ongoing
computation. Thus, HTBAC has to support resource redistribution, according to
the logic of the adaptive algorithms, enabling the optimization of
computational efficiency.

Finally, usability plays an important role in the development of HTBAC. HTBAC
has to provide a flexible interface which enables users to easily scale the
number of drug candidates and quickly prototype existing and novel free
energy protocols.

\subsection{Design and Implementation}\label{ssec:design_arch}

HTBAC exposes four constructs to specify free energy protocols: Protocol,
Simulation, Analysis, and Resource. \textbf{Protocol} enables multiple
descriptions of protocol types, while \textbf{Simulation} and
\textbf{Analysis} specify simulation and analysis parameters for each
protocol. \textbf{Resource} allows to specify the amount of resources needed
to execute the given protocols. Together, protocol instances, simulation and
analysis parameters, and resource requirements constitute an HTBAC
\textbf{application}.

Each protocol models a unique protein ligand physical system. Protocols
follow a sequence of simulation and analysis steps, assigning ensemble
members to execute independent simulations or analysis. An ensemble member
that executes a simulation within a simulation step is referred to as a
replica. Each simulation is assigned a different initial velocity, which
enables simulations to begin in different parts of the ligand's phase space.

Individual simulations or analyses with input, output, termination criteria
and dedicated resources are designed as a computational
\textbf{task}~\cite{power-of-many17}. Aggregates of tasks with dependencies
that determine the order of their execution constitute a \textbf{workflow}.
In this way, HTBAC encodes $N_P$ instances of the P$^{th}$ protocol as a
workflow of computational tasks.

\begin{figure}
  \centering
  \includegraphics[width=\columnwidth]{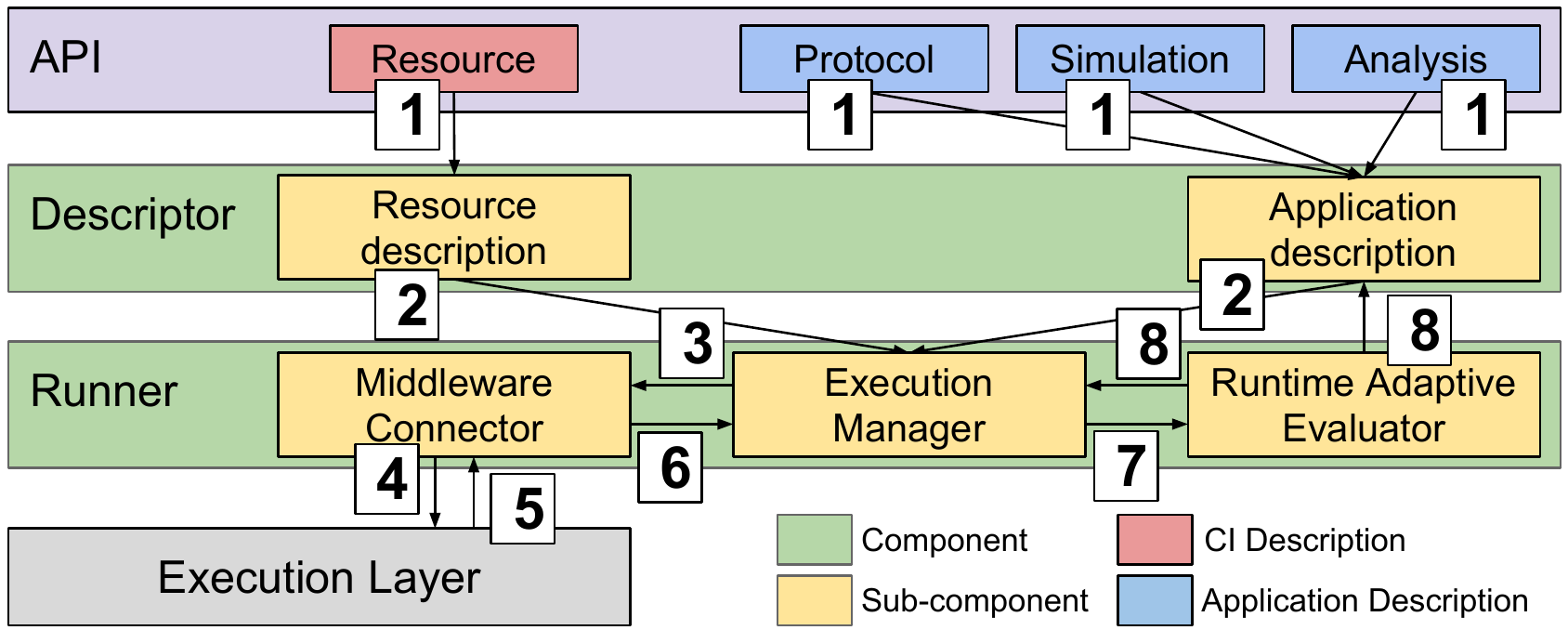}
  \caption{HTBAC architecture. Users specify protocol(s) with multiple
  simulation and analysis steps. Descriptor derives a single application that
  Runner executes on an external execution layer. Runtime Adaptive Evaluator
  enables the execution of adaptive protocols.}\label{fig:architecture}
\up{}
\up{}
\end{figure}

Fig.~\ref{fig:architecture} shows the components and subcomponents of HTBAC.
The API enables users to describe protocols in terms of protocol type,
simulation and analysis steps, and compute infrastructure requirements. The
Descriptor component uses two subcomponents to aggregate protocol
descriptions into a single application and resource description. Note that
Descriptor can aggregate different types of protocols, with different
computing and resource requirements.

The Runner component has three subcomponents: Execution Manager, Middleware
Connector and Runtime Adaptive Evaluator. Execution Manager communicates with
the execution layer via a connector to coordinate the execution of the
application. In principle, HTBAC can use multiple connectors for diverse
middleware to access different computing infrastructures.

Middleware Connector converts the application description of HTBAC into a
middleware-specific format. Execution Manager can pass the given application
to the connector in full or only in parts. This enables to start the
execution of an application before its full description is available or to
change those parts of the application that still have to be executed. This
will enable future capabilities like, for example, to concurrently execute
the application on diverse middleware.

Runtime Adaptive Evaluator enables the execution of adaptive applications.
This subcomponent can evaluate partial results of an application execution
via tailored algorithms. On the base of this evaluation, Runtime Adaptive
Evaluator can decide to return the control to Execution Manager or to modify
the description of the application that is being executed. In this way, HTBAC
implements adaptivity for diverse protocols, allowing users to define
arbitrary conditions and algorithms.

HTBAC is implemented in Python as a domain-specific library. All components
of HTBAC are implemented as objects that communicate via method calls. HTBAC
uses two RCT as building blocks~\cite{bb_ccgrid18}: Ensemble Toolkit
(EnTK) and RADICAL-Pilot (RP).

EnTK provides HTBAC capabilities to execute ensemble-based
applications~\cite{power-of-many17}. EnTK exposes three constructs:
\textbf{Task}, \textbf{Stage} and \textbf{Pipeline}. Tasks contain
information regarding an executable, its software environment and its data
dependencies. Stages are sets of tasks without mutual dependencies that can
execute concurrently. Pipelines are lists of stages, where stages can execute
only sequentially. Pipelines can execute independently. HTBAC uses a
Middleware Connector for EnTK to encode a protocol instance as a single
pipeline that contains stages of individual simulations and analyses tasks.

EnTK is designed to be coupled with different runtime systems. In this paper,
EnTK uses RP to execute tasks via pilots. RP supports task-level parallelism
and high-throughput by acquiring resources from a computing infrastructure
and scheduling tasks on those resources for execution. RP uses RADICAL-SAGA
to interface with several resource managers, including SLURM, PBS (pro), and
LSF. Pilot systems execute tasks directly on the resources, without queuing
them on the infrastructure's scheduler.

\subsection{Execution Model}\label{ssec:adaptive_execution}

Users describe one or more protocols alongside their
resource requirements via HTBAC's API. Descriptor takes these descriptions as
input and returns an application description (Fig.~\ref{fig:architecture}.1).
As seen in \S\ref{ssec:design_arch}, this application consists of a set or
sequence of tasks with a set of resource requirements for their execution.

The application description is passed to the Execution Manager of the Runner
component (Fig.~\ref{fig:architecture}.2). Execution Manager evaluates the
resource requirements, selects a suitable connector (currently only to EnTK),
tags each protocol instance of the application with an ID, and passes all or
part of the application description to the connector for execution
(Fig.~\ref{fig:architecture}.3).

The Middleware Connector of the Runner component gets the application
description, converting it into a middleware-specific description (EnTK
pipelines of stages of tasks) and a resource request. The connector submits
this request to the underlying execution layer
(Fig.~\ref{fig:architecture}.4) and initiates the execution of the
application once the execution layer communicates the availability of the
resources (Fig.~\ref{fig:architecture}.5).

The resource requirements specified via HTBAC's API include walltime, cores,
queue, and user credentials. EnTK derives a resource request from these
requirements, converting it into a pilot description for RP. RP converts this
pilot requests into a batch script that is submitted to the specified HPC
machine. Once the pilot becomes active, EnTK identifies those application
tasks that have satisfied dependencies and can be executed concurrently.
EnTK's own Execution Manager uses RP to execute those tasks on pilot's
resources.

HTBAC allows to specify conditions tailored to individual simulation steps of
a protocol implementation. We leverage this ability to implement adaptivity
by enabling the user to partition protocols into simulation steps and
generate new simulation steps at runtime, based on a set of predefined
conditions. The user specifies these conditions in an analysis script for the
Runtime Adaptive Evaluator subcomponent.

Execution Manager can retrieve the results of simulations
(Fig.~\ref{fig:architecture}.6) and these results can be evaluated by Runtime
Adaptive Evaluator via a user-defined analysis script
(Fig.~\ref{fig:architecture}.7). Depending on the result of the evaluation,
Runtime Adaptive Evaluator may generate new simulation steps, adding them to
the application description (Fig.~\ref{fig:architecture}.8a) or return the
control to Application Manager (Fig.~\ref{fig:architecture}.8b) without
changing the application. If new simulations are to be generated, the
Execution Manager bypasses termination of the application, and passes the
added application description to the connector.


In an adaptive scenario, as the number of simulations grows at runtime, the
ratio of cores-to-task fluctuates. EnTK's Execution Manager automatically
redistributes an even share of the total requested cores to each simulation.
RP allows for new simulations to execute within the pilot's wall-time,
without having to acquire new resources via the resource management system.

\subsection{Implementing ESMACS and TIES in
HTBAC}\label{sec:implementation_htbac}

In \S\ref{ssec:esm_ties} we define the structure of the ESMACS and TIES
protocols. Here we provide skeletons of the TIES protocol implemented in
HTBAC\@. In L.~\ref{lst:ties.py} we show a customization of a production
MD simulation step.

\begin{lstlisting}[language=Python, label={lst:ties.py}, caption={TIES protocol
implemented with HTBAC. We import the predefined protocol `TIES'. We assign
the physical system to the protocol, we instantiate a simulation, customize
its steps (\texttt{replica}, \texttt{lambda}) and assign it to the TIES's
\texttt{step0}. We instantiate Runner with a resource request and pass the
protocol to it.}]
import htbac.protocols.TIES
ties_1 = Protocol(system = 'brd4-1')
sim = Simulation(name = 'Production MD')
        ... # define simulation conditions 
sim.ensemble('replica', range(5))
sim.ensemble('lambda', range(13))
# add simulation configurations to protocol
TIES.step0 = sim
ties_1.append(TIES.step0)
# assign resources, append protocol to Runner 
runner = Runner(resource = 'ncsa.blue_waters',
                walltime = 60) 
runner.add_protocol(protocol = ties_1)
# launch application
runner.run()
\end{lstlisting}

In \S\ref{ssec:adaptive_execution} we show HTBAC's adaptive execution
capabilities. In L.~\ref{lst:ties_adaptivity.py} we provide an
intra-protocol adaptive implementation of TIES, based on the use-case
of \S\ref{ssec:adapt_ties}.

\begin{lstlisting}[language=Python, label={lst:ties_adaptivity.py},
caption={Adaptive TIES protocol implemented with HTBAC. Assuming
L.~\ref{lst:ties.py}, we run Runner retrieving runtime results, we specify an
adaptivity script for the evaluator, create TIES's \texttt{step1}. The
analysis script operates on partial simulation results, generating new
simulation conditions for the next simulation step.}]
# we start with the same previous implementation
# provide runner with two flags 
runner.run(save_output = True, terminate = False)
# specify adaptivity script 
requested_lambdas = AdaptiveQuadrature()
sim1 = Simulation(name = 'production MD 2')
        ... 
sim1.ensemble('replica', range(5))
sim1.ensemble('additional lambdas', 
    requested_lambdas)
TIES.step1 = sim1
ties_1.append(TIES.step1)
runner.add_protocol(protocol = ties_1)
runner.run(terminate = True)
\end{lstlisting}

%% file: experiments.tex
Typically, a computational campaign for drug discovery explores a large
number of drug candidates by running several workflows multiple times, each
requiring thousands of concurrent simulations. Before embarking on a campaign
that will utilize 150 million core-hours on NCSA Blue Waters, we perform
experiments to characterize the weak and strong scaling performance of HTBAC
and its overheads on Blue Waters. We validate the results of the free energy
calculations produced using HTBAC against published results.

Given that protocols like TIES are more computationally demanding than
protocols like ESMACS, it is paramount to use resources efficiently,
especially for campaigns that have a predefined computational budget. As
described in \S~\ref{sec:science-drivers} and~\ref{sec:htbac}, adaptive
simulation methods have the potential to reduce the number of simulations
without a loss in accuracy and with a lower computational load. We run
experiments with an adaptive implementation of TIES in HTBAC, measuring the
benefits in terms of accuracy, reduced number of simulations and
computational load.

\subsection{Experiment Setup}\label{ssec:exp_design}

Table~\ref{tab:experiments} shows 9 experiments we designed to characterize
the behavior of HTBAC on Blue Waters. Each experiment executes the ESMACS
and/or TIES protocol for different physical systems. Experiments 1--6 use the
BRD4 physical system provided by GlaxoSmithKline, while experiments 7--9
utilize the PTP1B, MC1, and TYK2 physical systems.

\begin{table*}
    \caption{Parameters of scalability and adaptivity
    experiments.}\label{tab:experiments}
    \centering
    \begin{tabular}{l                    
                    l                    
                    l                    
                    l                    
                    l                    
                    l                    
                    }
    \toprule
    \B{ID}                            &  
    \B{Type of Experiment}            &  
    \B{Physical System(s)}            &  
    \B{Protocol(s)}                   &  
    \B{No. Protocol(s)}               &  
    \B{Total Cores}                   \\ 
    \midrule
    \B{1}                             &  
    Weak scaling                      &  
    BRD4                              &  
    ESMACS                            &  
    (2, 4, 8, 16)                     &  
    1600, 3200, 6400                  \\ 
    \B{2}                             &  
    Weak scaling                      &  
    BRD4                              &  
    TIES                              &  
    (2, 4, 8)                         &  
    4160, 8320, 16640                 \\ 
    \B{3}                             &  
    Weak scaling                      &  
    BRD4                              &  
    ESMACS + TIES                     &  
    (2, 4, 8)                         &  
    5280, 10560, 21120                \\ 
    \B{4}                             &  
    Strong scaling                    &  
    BRD4                              &  
    TIES                              &  
    (8, 8, 8)                         &  
    16640, 8320, 4160                 \\ 
    \B{5}                             &  
    Strong scaling                    &  
    BRD4                              &  
    ESMACS                            &  
    (16, 16, 16)                      &  
    6400, 3200, 1600                  \\ 
    \B{6}                             &  
    Strong scaling                    &  
    BRD4                              &  
    ESMACS + TIES                     &  
    (20, 20, 20)                      &  
    22120, 10560, 5280                \\ 
    \B{7}                             &  
    Non-adaptivity                    &  
    PTP1B, MC1, TYK2                  &  
    TIES                              &  
    (1, 1, 1)                         &  
    2080, 2080, 2080                  \\ 
    \B{8}                             &  
    Adaptivity                        &  
    PTP1B, MC1, TYK2                  &  
    TIES                              &  
    (1, 1, 1)                         &  
    2080, 2080, 2080                  \\ 
    \B{9}                             &  
    Reference                         &  
    PTP1B, MC1, TYK2                  &  
    TIES                              &  
    (1, 1, 1)                         &  
    10400, 10400, 10400               \\ 
    \bottomrule
    \end{tabular}
\up{}
\end{table*}

Experiment 1 and 2 measure the weak scaling of HTBAC using the ESMACS and
TIES protocols. Experiments 3 uses both the TIES and ESMACS protocols,
characterizing the weak scaling of heterogeneous protocol executions.
Experiments 4 and 5 measure the strong scaling of HTBAC using a fix number of
instances of the ESMACS and TIES protocols. Experiments 6 uses both the TIES
and ESMACS protocols, characterizing the strong scaling of heterogeneous
protocol executions. Experiments 7--9 characterize nonadaptive and adaptive
simulation methods using the TIES protocol.

In each weak scaling experiment (1--3), we keep the ratio between resources
allocated and protocol instances constant. Consistently, for each experiment
we progressively increase both the number of cores (i.e., measure of
resource) and the number of protocol instances by a factor of 2. In each
strong scaling experiment (4--6), we change the ratio between resources
allocated and the number of protocol instances: we fix the number of protocol
instances and reduce the number of cores by a factor of 2.

Weak scaling experiments provide insight into the size of the workload that
can be executed in a given amount of time, while strong scaling experiments
show how the time duration of the workload scales when adding resources. For
all the weak and strong scaling experiments we characterize the overheads of
HTBAC, EnTK and RP, and we show an approximation of the time taken by the
resources to become available. This offers insight about the impact of HTBAC
and its runtime system on the time to completion of each workload.
In~\cite{dakka2017}, we show baseline performance of HTBAC using ESMACS with
a null workload.

For weak and strong scaling experiments, we reduced the number of time-steps
of the protocols and omitted the analysis steps $S5$ and $S6$ of their
workflows (Fig.~\ref{fig:ties_esmacs_application}). These simplifications are
consistent with characterizing scalability performance instead of simulation
duration. The time-steps are set to enable the physical systems to reach 
steady-state. For the experiments 1--6 we used the following 
time-steps: $S1=1000$; $S2=5000$; $S3=5000$; and $S4=50000$.

We measure the following durations for Experiments 1--6:
\begin{itemize}
    \item \textbf{Total Task Execution Time}: Time taken by all the task
    executables to run on the computing infrastructure.
    \item \textbf{HTBAC Overhead}: Time taken to instantiate HTBAC, and
    validate and process the application description.
    \item \textbf{EnTK and RP Overhead}: Time taken by EnTK and RP to manage
    the execution of tasks.
    \item \textbf{\texttt{aprun} Overhead}: Time taken by \texttt{aprun} to
    launch tasks on Blue Waters.
\end{itemize}

Note that once RP relinquishes the control flow to \texttt{aprun}, the
precise time at which \texttt{aprun} schedules each task on a compute node
and the MD kernel of each task begins execution cannot be measured. Instead,
for each task, we measure the difference between the task execution time and
its \texttt{NAMD} kernel execution time, provided by the \texttt{NAMD} output
logs. In this way, we approximate the time taken by \texttt{aprun} to launch
the task. Once aggregated, these measures constitute what we defined as
\texttt{aprun} Overhead. The summation of all durations provides the average
wall-time of the pilot job.

Experiments 7--9 compare the accuracy and time to solution of nonadaptive and
adaptive simulation methods. For the nonadaptive simulation method of
Experiment 7 we use 13 preassigned and approximated $\lambda$ windows,
consistent with the value reported in Ref.~\cite{Bhati2017}. In this way, we
produce 65 concurrent simulations for stages $S1$--$S4$ of TIES (see
Fig.~\ref{fig:ties_esmacs_application}). The production simulation stage $S4$
executes each simulation for \SI{4}{\nano\second}. Stage $S5$ has 5 analysis
tasks which aggregate the simulation results of $S4$. The global analysis
stage $S6$ has a single task that aggregates the results from $S5$.

\begin{figure}
  \centering
  \includegraphics[width=\columnwidth]{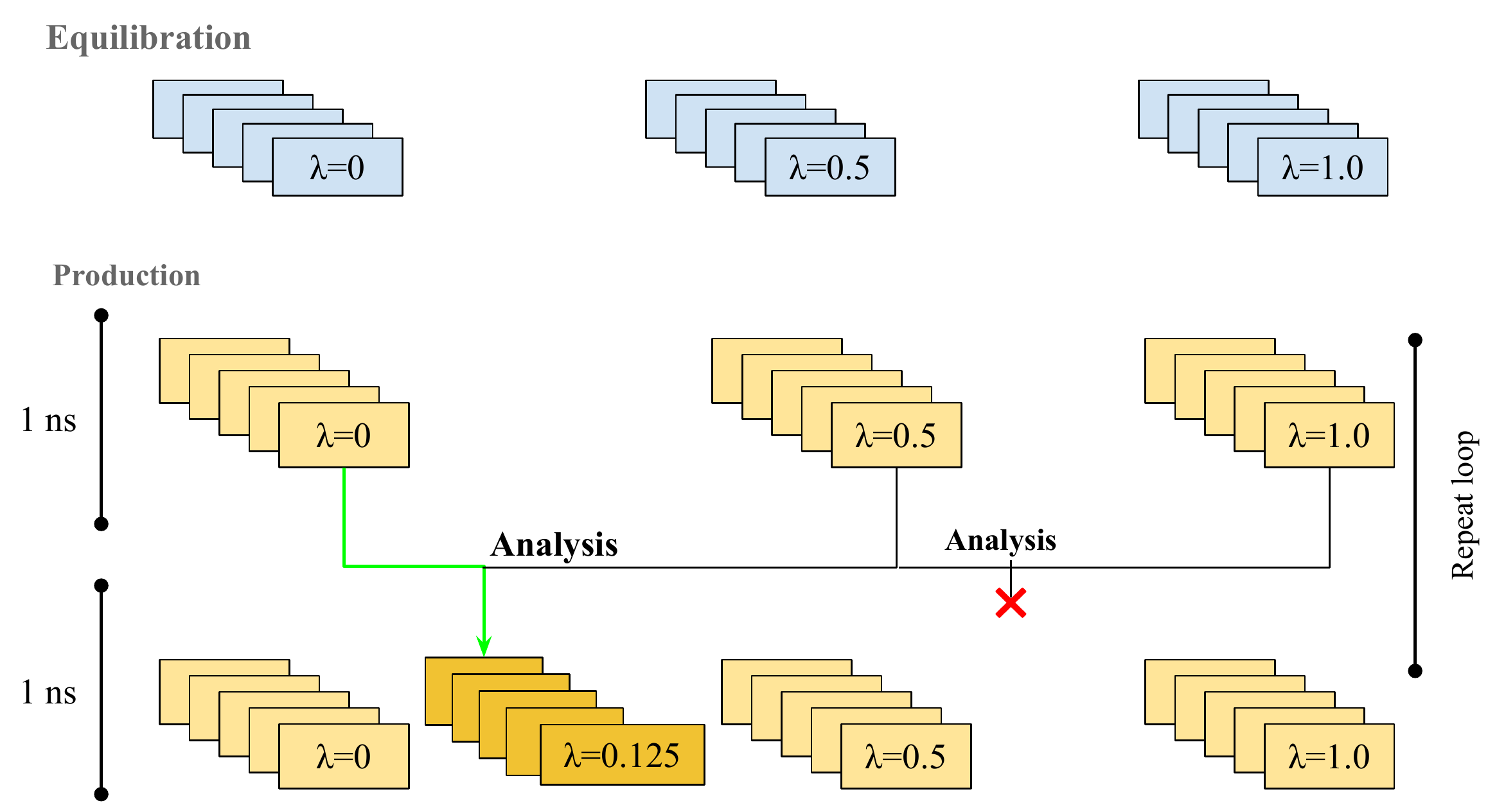}
  \caption{Adaptive workflow for TIES. After equilibrating 3 $\lambda$, the
  first stage starts. This is followed by analysis at every $\lambda$
  interval, to decide whether to add a new window in between. In our
  implementation, the simulation-analysis cycle is repeated for 4 simulation
  steps, not shown here.}\label{fig:adaptive_TIES}
\up{}
\up{}
\up{}
\end{figure}

In the adaptive implementation~\ref{fig:adaptive_TIES}, we initialize the
TIES protocol with 3 $\lambda$ windows, obtaining 15 replicas. We separate
stage $S4$ of each TIES replica into 4 sub-stages. Each sub-stage runs a
\SI{1}{\nano\second} simulation, followed by an adaptive quadratures analysis
which estimates free energy errors with respect to each interval of two
$\lambda$ values.

We use Experiment 9 to compare the adaptive and non-adaptive execution of
TIES. We use 
65 simulations, derived from 13 equally spaced $\lambda$ windows to calculate
the free energy with high accuracy. This creates a baseline against which to
compare the adaptive and non-adaptive results.

We assigned the following simulation time-steps in Experiment 7 and 9:
$S1=3000$; $S2=50000$; $S3=50000$; and $S4=2000000$. The adaptive simulation
of Experiment 8 uses the same time-steps, apart from $S4$ which is divided
into 4 sub-stages of 500000 time-steps each.

We performed all the experiments on Blue Waters, a 26868 node Cray XE6/XK6
SuperComputer with peak performance of 13.3 petaFLOPS managed by NCSA.
Consistent with NCSA policies, we initiated the experiments from a virtual
machine outside NCSA, avoiding to run persistent process on the NCSA login
node. We used HTBAC 0.1, EnTK 0.6, and RP 0.47 and the \texttt{NAMD-MPI} MD
kernel, and launched via the \texttt{aprun} command. For the analysis stages
in the TIES protocol we used \texttt{AmberTools}.

NCSA sets a system policy on the maximum number of processes that
\texttt{aprun} can spawn, limiting the number of concurrent tasks we can
execute on Blue Waters to $\approx$450. During the execution of Experiment 2,
we observed failing tasks with 8 TIES protocol instances, i.e., 520
concurrent tasks. In a trial of 10 repetitions at this scale, we observed an
average of $70\pm6.67$ failing tasks. More data would be required to model
the distribution type of these results.

NCSA allows to run only one MPI application for each compute node. Thus, we
run each MD simulation with 32 cores (i.e., one compute node) even if our
performance of \texttt{NAMD} on Blue Waters indicated that 16 cores offers
the best trade-off between computing time and communication overhead.

\subsection{Weak Scaling Characterization}

\begin{figure}
  \centering
    \includegraphics[width=\columnwidth]{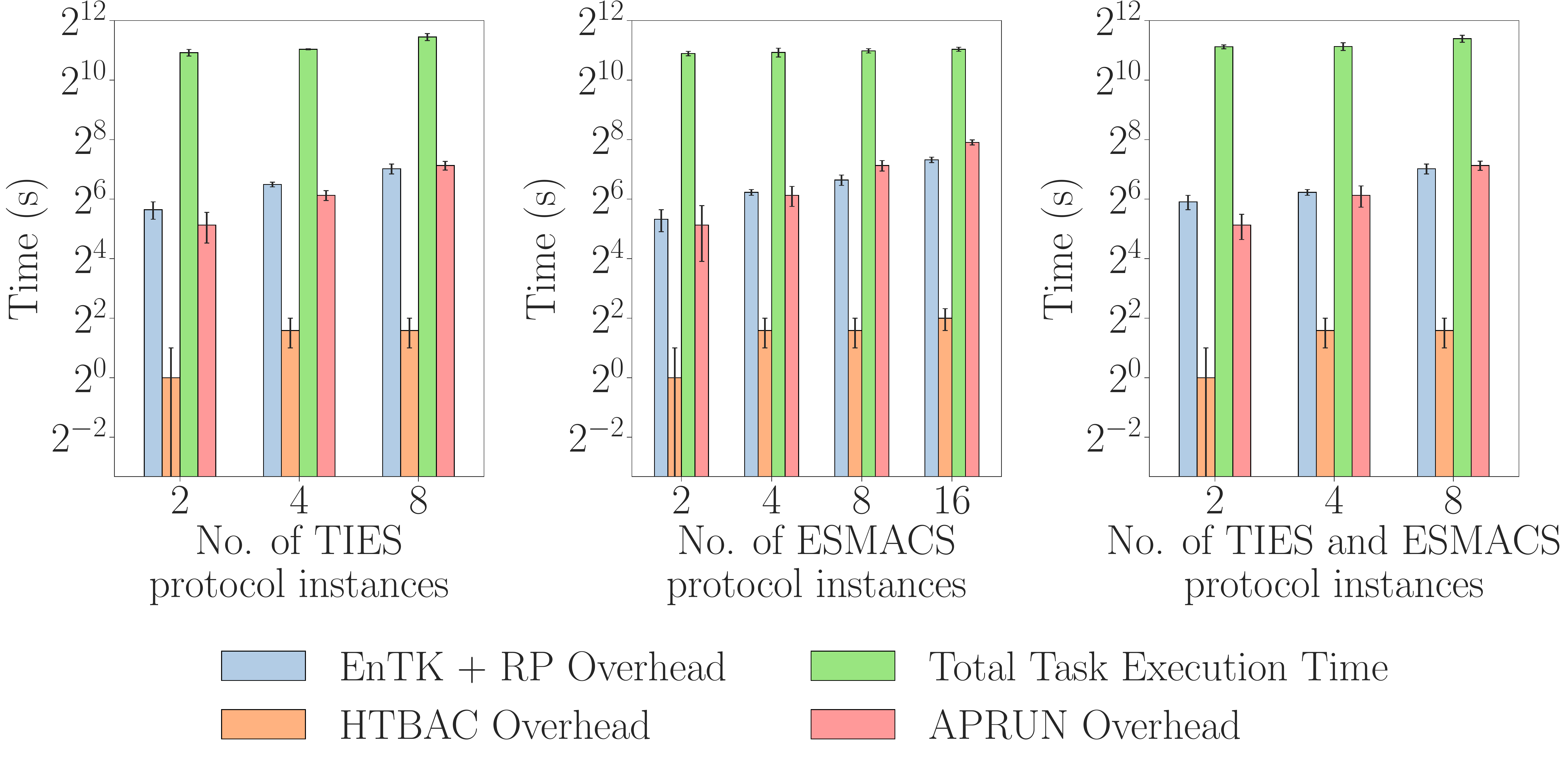}
    \caption{Weak scaling of HTBAC. The ratio number of protocol instances to
    resources is constant. Task Execution Time, and HTBAC, EnTK+RP,
    \texttt{aprun} overheads with (a) TIES (Experiment 1), (b) ESMACS
    (Experiment 2), and (c) TIES and ESMACS (Experiment 3).}\label{fig:ws}
\up{}
\up{}
\up{}
\end{figure}

Fig.~\ref{fig:ws}(a) shows the weak scaling of HTBAC with the TIES protocol.
Each instance of the TIES protocol contains a single pipeline with 4 stages
and 65 concurrent tasks. We increase the number of protocol instances
linearly, between 2 and 8. When scaling to 8 protocol instances, we execute
more than 450 concurrent tasks, the average limit supported by
\texttt{aprun}, as described in \S\ref{ssec:exp_design}. This introduces some
failures that contribute to a slight degradation in performance.

Fig.~\ref{fig:ws}(b) shows the weak scaling of HTBAC with the ESMACS
protocol. We increase the number of instances linearly, between 2 and 16.
Each ESMACS protocol contains 1 pipeline with 4 stages and 25 concurrent
tasks.

Fig.~\ref{fig:ws}(c) shows the weak scaling of HTBAC with instances of both
TIES and ESMACS protocols. Also in this case, we scale the instances of both
protocols linearly, between 2 and 8. The first configuration shows 1 ESMACS
and 1 TIES protocol, and with each increase in scale we double the number of
protocols. Experiments 2 and 3 show scaling ranges within the limit of the
maximum number of concurrent tasks we can successfully execute on Blue
Waters.

For all weak scaling experiments (1--3) we use physical systems from the
\texttt{BRD4-GSK} library with the same number of atoms and similar chemical
properties. The uniformity of these physical systems ensures a consistent
workload with insignificant variability when characterizing their performance
under different conditions.

In all weak scaling experiments (Fig.~\ref{fig:ws}) we observe that the value
of Total Task Execution Time (green bar) shows minimal variation as the
number of protocol instances increases, suggesting that HTBAC is invariant to
the protocol. We conclude that HTBAC shows near-ideal weak scaling behavior
under these conditions.

The HTBAC overhead depends mostly on the number of protocol instances that
need to be generated for an application. This overhead shows a super linear
increase as we grow the number of protocol instances, but the duration of the
overhead is negligible when compared to Total Task Execution Time.

The \texttt{aprun} overhead increases as we approach the limit of concurrent
\texttt{aprun} processes that can be executed on Blue Waters. For example,
when scaling to 8 TIES protocol instances (Fig.~\ref{fig:ws}(a)), we see that
the increase in \texttt{aprun} overhead occurs due to task failure. This is
explained by noticing that attempts to relaunch failed tasks require
additional communication among the nodes that were running the tasks and the
MOM Nodes from which the execution is coordinated.

EnTK and RP overheads mostly depend on the number of tasks that need to be
translated in-memory from a Python object to a task
description~\cite{dakka2017,merzky2018}. As such, those overheads are
expected to grow proportionally to the number of tasks, as observed in
Fig.~\ref{fig:ws}, blue bars.

The RP overhead is calculated by measuring and aggregating the execution time
of the RP components that manage and coordinate the execution of the protocol
instances. Among these components, the task scheduler of RP introduces the
largest overhead. Due to the general scheduling algorithm loaded by default
in RP, the task scheduling overhead scales linearly with the number of tasks
that need to be scheduled.

In comparison to Total Task Execution Time, the EnTK and RP overheads are an
order of magnitude shorter, yet they directly contribute to the total
duration of the application execution. Based on Fig.~\ref{fig:ws}, we
approximate the use of our systems will results in $\approx15\%$ additional
usage of resource allocation. This overhead can be substantially reduced by
using a special-purpose scheduler for RP as illustrated in
Ref.~\cite{merzky2018}.

\subsection{Strong Scaling Characterization}

In Experiment 4 we fix the number of instances of the TIES protocol to 8 (due
to the described \texttt{aprun} limitations) and we vary the amount of
resources between 4160, 8320 and 16640 cores. Assuming the definition of
`generation' in \S\ref{ssec:exp_design}, given 4160 cores, we can execute 4
generations of 130 concurrent tasks; with 8320 cores, 2 generations of 260
tasks; and with 16640 cores, 1 generation of 520 tasks.

In Experiment 5 we fix the number of instances of the ESMACS protocols to 16
and vary the amount of resources between 3200, 6400 and 12800 cores. In this
way, we obtain the same number of generations as in Experiment 4.

In Experiment 6 we fix the number of instances of the ESMACS and TIES
protocols to 16 and 4 respectively, and vary the amount of resources between
5280, 10560 and 22120 cores. In this way, we obtain the same number of
generations as in Experiment 4 and 5.

\begin{figure}
  \centering
   \includegraphics[width=\columnwidth]{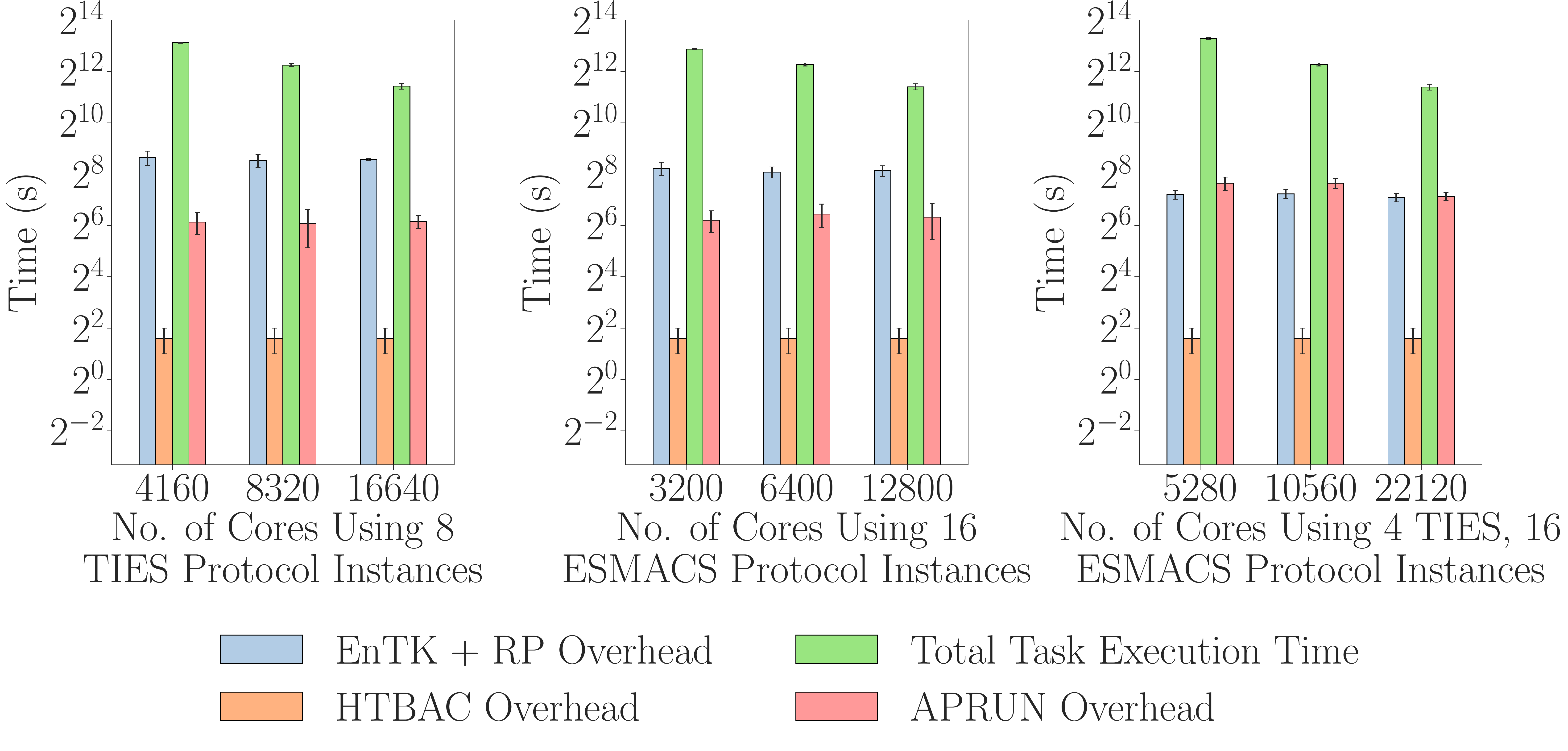}
   \caption{Strong scaling of HTBAC. The number of protocol instances is
    fixed while the number of cores increases. Task Execution Time, and
    HTBAC, EnTK+RP, \texttt{aprun} overheads with (a) TIES (Experiment 4),
    ESMACS (Experiment 5) and ESMACS + TIES (Experiment 6).}\label{fig:ss}
\up{}
\up{}
\up{}
\end{figure}

Fig.~\ref{fig:ss} shows a linear speedup in Total Task Execution Time for
both experiments, proportional to the increase in the number of cores. The
availability of more resources for a fixed number of protocols explains this
behavior. Overheads remain essentially constant for both experiments when
increasing the number of cores. The scheduling of the number of tasks, as
opposed to the amount of resources, is the main driver of EnTK and RP
overheads (Ref.~\cite{merzky2018}).

\subsection{Validation}

In order to validate the correctness of the results produced in Experiment
1--6, using HTBAC and the BRD4-GSK physical systems, we compare our results
with those previously published in Wan et al.~\cite{Wan2017brd4}. In this
way, we can confirm that we calculated the correct binding free energies
values.

We validated our implementation selecting a subset of the protein ligand
systems used in Wan et al.~\cite{Wan2017brd4}: ligand transformations 3 to 1,
4, and 7. We then performed a full simulation on all 3 systems and calculated
the binding affinity using HTBAC.

The results of our experiments, collected in Table~\ref{tab:exp2}, show that
all three $\Delta \Delta$G values are within error bars of the original
study, validating the results we produced with HTBAC.

\begin{table}
  \centering
  \caption{Validation of HTBAC results against published and experimental
  values}\label{tab:exp2}  
  \begin{tabular}{lrrr}
    \toprule
    System & 
    {\makecell{HTBAC \\ (\si{\kilo\calorie\per\mole})}} & 
    {\makecell{Wan et al. \\ (\si{\kilo\calorie\per\mole})}} & 
    {\makecell{Experiment \\ (\si{\kilo\calorie\per\mole})}} \\
    \midrule
    BRD4 \textbf{3 to 1} & \num{0.39 +- 0.10} &   \num{0.41 +- 0.04} &  \num{0.3 +- 0.09} \\
    BRD4 \textbf{3 to 4} & \num{0.02 +- 0.12} &   \num{0.01 +- 0.06} &  \num{0.0 +- 0.13} \\
    BRD4 \textbf{3 to 7} & \num{-0.88 +- 0.17} &  \num{-0.90 +- 0.08} & \num{-1.3 +- 0.11} \\
    \bottomrule
  \end{tabular}
\up{}
\up{}
\up{}
\end{table}

\subsection{Adaptive Experiments}

The design of HTBAC permits enhancing protocols while continuing to use
``static'' simulation engines. To this end, we implemented two adaptive
methods using HTBAC: adaptive quadrature and adaptive termination. Both of
these methods use the features of adaptivity offered in HTBAC to scale to
large number of concurrent simulations and to increase convergence rate and
obtain more accurate scientific results.

The aim of introducing adaptive quadrature for alchemical free energy
calculation protocols (e.g., TIES) is to reduce time to completion while
maintaining (or increasing) the accuracy of the results. Time to completion
is measured by the number of core-hours consumed by the simulations. Accuracy
is defined as the error with respect to a reference value, calculated via a
dense $\lambda$ window spacing (65 windows). This reference value is used to
establish the accuracy of the non-adaptive protocol (which has 13 $\lambda$
windows) and the adaptive protocol (which has a variable number of $\lambda$
windows, determined at run time).

One of the input parameters of the adaptive quadrature algorithm is the
desired acceptable error threshold of the estimated integral. We set this
threshold to the error of the non-adaptive algorithm calculated via the
reference value. The algorithm then tries to minimize the number of $\lambda$
windows constrained by the accuracy requirement.

\begin{figure}
  \input{fig-savings.tex}%
  \caption{Quantifying the benefits of the adaptive quadrature simulations.
  (top) The error of the adaptive run is reduced for all 5 test systems,
  sometimes by a significant amount. It has been shown that reproducibility
  of free energy calculations can be achieved up to
  \SI{0.2}{\kilo\calorie\per\mole}~\cite{Loeffler2018}. The adaptive
  algorithm brings down the error of the nonadaptive simulations below this
  threshold, ensuring that results are also reproducible. (bottom) Resource
  consumption is reduced, except for one of the systems, where the low error
  threshold required more $\lambda$ windows.}\label{fig:savings}
\up{}
\up{}
\end{figure}
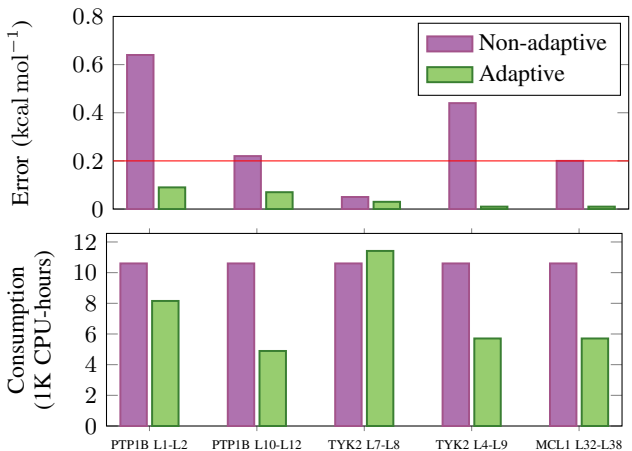

Table~\ref{tab:adapquad} shows the results of running adaptive quadrature on
5 protein ligand systems, comparing the Total Task Execution Time and
accuracy versus the non-adaptive case. The number of lambda windows are
reduced on average by \SI{32}{\percent}, hence reducing Total Task Execution
Time by the same amount. The error on the adaptive results is also decreased,
on average by \SI{77}{\percent} (see fig.~\ref{fig:savings}). More
importantly, the error on all of the systems are reduced to below
\SI{0.2}{\kilo\calorie\per\mole}, which has recently been shown to be the
upper bound of reproducibility across different simulation
engines~\cite{Loeffler2018}.

The Total Task Execution Time of the TYK2 L7--L8 system has increased for the
adaptive run by 1 $\lambda$ window, compared to the non-adaptive case. This
is due to the non-adaptive error being very low, and matching that same
accuracy required the use of a large number of windows. Nonetheless due to
the efficient placing of the windows, the accuracy of the free energy still
increased by \SI{40}{\percent}.

\begin{table*}
  \caption{Comparing results of adaptive, non-adaptive and reference
  runs.}\label{tab:adapquad}
  \begin{tabular}{lSSSSS[table-format=5.1]S[table-format=5.1]}
    \toprule
    {System}                               & 
    {\makecell{Ref $\Delta \Delta$G \\ (\si{\kilo\calorie\per\mole})}}  &
    {\makecell{Non-adaptive $\Delta \Delta$G \\ (\si{\kilo\calorie\per\mole})}}       &
    {\makecell{Adaptive $\Delta \Delta$G \\ (\si{\kilo\calorie\per\mole})}}           &
    {No. of $\lambda$ windows}            &
    {Decrease in TTX}       &
    {Increase in accuracy}                 \\
    \midrule
    {PTP1B L1-L2}   & 
    -58.51 & 
    -57.87(64) & 
    -58.60(9) & 
    10 & 
    23\si{\percent} & 
    86\si{\percent} \\
    {PTP1B L10-L12} & 
    1.83   & 
    2.05(22) & 
    1.94(7)  & 
    6  & 
    54\si{\percent} &
    68\si{\percent} \\
    {MCL1  L32-L38} & 
    2.13   & 
    2.33(20) & 
    2.14(1)      & 
    7  & 
    46\si{\percent} & 
    95\si{\percent} \\
    {TYK2  L4-L9}   &
    -28.69 & 
    -28.25(44) & 
    -28.67(1)  & 
    7  & 
    46\si{\percent} & 
    98\si{\percent} \\
    {TYK2  L7-L8}   & 
    4.97   & 
    4.92(5) & 
    5.00(3)      & 
    14 &  
    -8\si{\percent} & 
    40\si{\percent} \\
    \bottomrule 
  \end{tabular}
\end{table*}

\begin{figure}
  \input{fig-adapt.tex}
  \caption{Plot of the error estimate as a function of the resource
  consumption, comparing the adaptive and nonadaptive simulations. The error
  estimate converges for both simulations but the window placement of the
  adaptive simulation considerably lowered the error.}\label{fig:adapconv}
\up{}
\up{}
\end{figure}
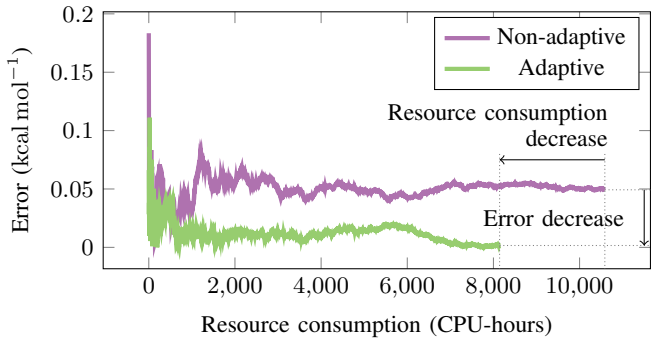

Fig.~\ref{fig:adapconv} compares the error on the adaptive and non-adaptive
simulations as a time series plot. As fewer lambda windows are calculated the
adaptive algorithm uses less resources. Remarkably, the error is drastically
reduced as the windows are placed adaptively to capture the changes in
function.

Adaptive quadrature is specific to alchemical free energy calculations.
\emph{Adaptive termination}, the second adaptive method implemented in HTBAC,
offers dynamic termination for any simulation protocol that has as its aim
the prediction of an observable value. The protocol monitors the convergence
of the observable as the simulation progresses, and stops the execution when
a criterion has been met. Non-adaptive protocols usually have a predefined
simulation time, set based on the assumption that the simulation will
converge by that time. This means that in practical examples the simulation
might have converged before the predefined simulation time.

\begin{figure}
  \input{fig-term.tex}
  \caption{Plot of the free energy as a function of the resources consumed
  (hence simulation time). The adaptive termination algorithm checks the
  convergence of the observable every $\tau = \SI{0.5}{\nano\second}$ and if
  the threshold (\SI{0.01}{\kilo\calorie\per\mole}) has been met, terminates
  the simulation.}\label{fig:termination}
\end{figure}
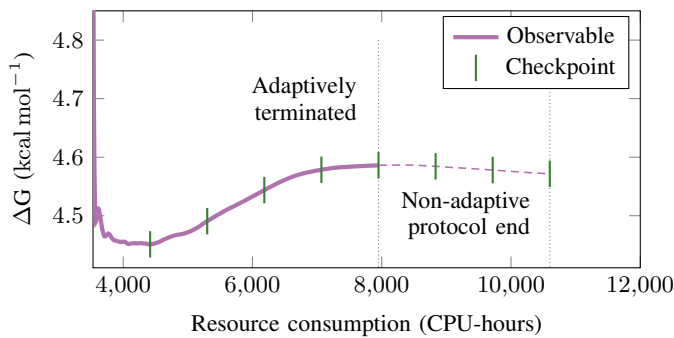

In the original TIES protocol the production part of the simulation has to be
run for \SI{4}{\nano\second} and the results are analyzed thereafter. This
assumes that all systems need this simulation time for the results to
converge. In reality, certain systems could converge faster, therefore one
can terminate the simulation before the static \SI{4}{\nano\second} end. This
would lead to faster time to insight and less compute resources consumed.
Adaptive termination was implemented in HTBAC by having a checkpoint every
$\tau = \SI{0.5}{\nano\second}$ in the simulation. Fig.~\ref{fig:termination}
shows how the observable for a specific simulation changes as a function of
resource consumption. At every checkpoint the convergence is evaluated, and
the simulation is indeed terminated earlier than the non-adaptive protocol
would suggest. Table~\ref{tab:adapterm} shows results that the adaptively
terminated TIES protocol saves compute resources and reduces time to insight
on average by \SI{16}{\percent} for the physical systems tested.

\begin{table}
  \caption{Simulation time of non-adaptive and adaptively terminated runs for
  a given convergence criterion}\label{tab:adapterm}
  \centering
  \begin{tabular}{lSSS[table-format=5.1, table-column-width=2cm]}
    \toprule
    {System}                               & 
    {Non-adaptive}      &
    {Adaptive}          &
    {Decrease in TTX}       \\
    \midrule
    {PTP1B L10-L12} & 
    6.0\si{\nano\second}   & 
    5.0\si{\nano\second}   & 
    16.7\si{\percent} \\
    {TYK2 L4-L9}   &
    6.0\si{\nano\second} & 
    5.5\si{\nano\second} & 
     8.3\si{\percent} \\
    {TYK2 L7-L8}   & 
    6.0\si{\nano\second}  & 
    4.5\si{\nano\second} & 
    25.0\si{\percent} \\
    \bottomrule 
  \end{tabular}
\up{}
\up{}
\end{table}

%% file: fig-savings.tex
\begin{flushleft}
\begin{tikzpicture}
\begin{axis}[
  font=\small,
  ybar,
  area legend,
  ymin=0,
  ymax=0.8,
  ylabel={Error (\si{\kilo\calorie\per\mole})\\},
  ylabel style={align=center},
  x tick label style  = {text width=1.5cm,align=center},
  xtick = \empty,
  symbolic x coords={PTP1B L1-L2,PTP1B L10-L12,TYK2 L7-L8,TYK2 L4-L9,MCL1 L32-L38},
  scaled ticks = false,
  yscale=0.45,
  ytick={0,0.2,0.4,0.6,0.8},
  legend cell align=left,
  legend style={at={(0.985,2.15)},anchor=north east}
  ]
  
\addplot+[draw=DarkOrchid, fill=Orchid, line width=0.75pt] table [x=System, y=Non-adaptive error, col sep=comma] {savings.csv};

\addplot+[draw=OliveGreen, fill=YellowGreen, line width=0.75pt] table [x=System, y=Adaptive error, col sep=comma] {savings.csv};

\coordinate (A) at (axis cs:PTP1B L1-L2,0.2);
\coordinate (O1) at (rel axis cs:0,0);
\coordinate (O2) at (rel axis cs:1,0);

\draw [red,sharp plot] (A -| O1) -- (A -| O2);

\legend{Non-adaptive,Adaptive} 
  
\end{axis}
\end{tikzpicture}
\begin{tikzpicture}
\begin{axis}[
  font=\small,
  x tick label style={font=\tiny},
  ybar,
  ymin=0,
  ylabel={Consumption\\(1K CPU-hours)},
  ylabel style={align=center},
  x tick label style  = {text width=1.5cm,align=center},
  symbolic x coords={PTP1B L1-L2,PTP1B L10-L12,TYK2 L7-L8,TYK2 L4-L9,MCL1 L32-L38},
  yscale=0.45,
  ]
  
\addplot+[draw=DarkOrchid, fill=Orchid, line width=0.75pt] table [x=System, y=Non-adaptive cpuh, col sep=comma] {savings.csv};

\addplot+[draw=OliveGreen, fill=YellowGreen, line width=0.75pt] table [x=System, y=Adaptive cpuh, col sep=comma] {savings.csv};

  
\end{axis}
\end{tikzpicture}%
\end{flushleft}%

%% file: fig-adapt.tex
\begin{tikzpicture}
\begin{axis}[
  height=5cm,
  width=\columnwidth,
  font=\small,
  no markers,
  every axis plot/.append style={ultra thick},
  xlabel=Resource consumption (CPU-hours),
  ylabel=Error (\si{\kilo\calorie\per\mole}),
  scaled ticks=false,
  yticklabel style={
  /pgf/number format/precision=3,
  /pgf/number format/fixed},
  ]

\addplot+[color=Orchid, smooth] table [x=Resource consumption, y=Non-adaptive error, col sep=comma] {non_adaptive_accuracy.csv};

\addplot+[color=YellowGreen, smooth] table [x=Resource consumption, y=Adaptive error, col sep=comma] {adaptive_accuracy.csv};

\legend{Non-adaptive,Adaptive}

\draw[densely dotted, color=gray] (10586.058921984,0.049299384672436553) -- (12000,0.049299384672436553);
\draw[densely dotted, color=gray] (8143.1222476800012,0.0015858767437348931) -- (12000,0.0015858767437348931);

\draw[densely dotted, color=gray] (10586.058921984,0.08) -- (10586.058921984,-0.025);
\draw[densely dotted, color=gray] (8143.1222476800012,0.08) -- (8143.1222476800012,-0.025);
\draw[->] (10586.058921984,0.075) -- (8143.1222476800012,0.075) ;
\node[align = right] at (8100, 0.105) {Resource consumption\\decrease};

\node at (9400, 0.025) {\contour{white}{Error decrease}};
\draw[->] (11500,0.049299384672436553) -- (11500,0.0015858767437348931) ;

\end{axis}
\end{tikzpicture}%

%% file: fig-term.tex
\begin{tikzpicture}
\begin{axis}[
  font=\small,
  height=5cm,
  width=\columnwidth,
  every axis plot/.append style={ultra thick},
  xlabel=Resource consumption (CPU-hours),
  ylabel=$\Delta$G (\si{\kilo\calorie\per\mole}),
  scaled ticks=false,
  yticklabel style={
  /pgf/number format/precision=3,
  /pgf/number format/fixed},
  xmin=3533.3975040000005,
  xmax=12000,
  ymax=4.85,
  ]

\addplot+[no markers, color=Orchid, smooth, error bars/.cd, y dir=both,y explicit, error mark=none, error bar style={draw=gray}] table [x=Resource consumption, y=dG, col sep=comma] {adaptive-termination-begin.csv};

\addplot [only marks, mark=|, mark options={color = OliveGreen, mark size = 5pt, thick}] coordinates {(4416.746880000001,4.451008938594887) (5300.096256000001,4.490752904442608) (6183.445632000001,4.543730747316334) (7066.795008000001,4.5784909284966355) (7950.144384000001,4.586120549368642) (8833.493760000001,4.5843340530391075) (9716.843136000001,4.577994475375011) (10600.192512000001,4.571614686607193) };

\addplot+[densely dashed, semithick, no markers, color=Orchid, smooth, error bars/.cd, y dir=both,y explicit, error mark=none, error bar style={draw=gray}] table [x=Resource consumption, y=dG, col sep=comma] {adaptive-termination-end.csv};

\legend{Observable, Checkpoint}

%
\draw[densely dotted, color=gray] (7950.144384000001,4.8) -- (7950.144384000001,4.4);
%
\node[align = right] at (6800, 4.7) {Adaptively\\terminated};

\draw[densely dotted, color=gray] (10600.192512000001,4.8) -- (10600.192512000001,4.4);

\node[align = right] at (9300, 4.5) {Non-adaptive\\protocol end};

%
  
\end{axis}
\end{tikzpicture}%

%% file: conclusions.tex
Ensemble-based binding affinity protocols have considerable predictive
potential in computational drug campaigns. As drug screening can cover
millions of compounds and hundreds of millions of core-hours, it is important
for binding affinity calculations to optimize the accuracy and precision of
results. However, the optimal protocol configuration for a given compound is
difficult to determine \textit{a priori}, thus requiring runtime adaptations
to workflow executions. We introduce HTBAC to enable scalable and adaptive
binding affinity energy calculations on HPC.

Specifically, this paper makes the following contributions: (1) shows how
adaptive execution of ensemble-based free energy protocol (TIES) improve
binding affinity accuracy given a fixed amount of computing resources; (2)
characterizes HTBAC, the software system we developed to enable the adaptive
execution of ensemble-based binding affinity protocols on HPC; and (3) shows
the capability to execute adaptive applications at scale, validating their
scientific results.

We characterize the performance of HTBAC on NCSA Blue Waters. We show
near-ideal weak and strong scaling behavior for ESMACS and TIES, individually
and together, reaching scales of 21,120 cores. Furthermore, we validate
binding free energies computed using HTBAC with both experimental and
previously published computational results.

We compare resource consumption and free energy accuracy in our adaptive and
non-adaptive TIES results. Using the adaptive quadrature algorithm, we show
improvements in $\Delta \Delta$G on average by 77\% over the 5 physical
systems tested. By reducing the $\lambda$ windows on average by 32\%, we
reduce execution time by the same amount. The adaptive termination
implementation of the TIES protocol saves compute resources and reduces time
to solution on average by 16\%. To the best of our knowledge, adaptive TIES
protocols have not been benchmarked against non-adaptive implementations
before.